\documentclass[12pt,a4paper]{article}
\usepackage{graphicx}
\usepackage{epsfig}
\usepackage{subfig}
\usepackage{amssymb}
\begin{document}

\title {Effects of maximal fluctuation moment $q$  and detrending polynomial orders on the observed multifractal features within MFDFA}

\author {Grzegorz Pamu{\l}a\footnote{gpamula@ift.uni.wroc.pl} and Dariusz Grech\footnote{dgrech@ift.uni.wroc.pl} \\
Institute of Theoretical Physics\\
Pl.
M. Borna 9, University of Wroc{\l}aw, PL-50-204 Wroc{\l}aw, Poland}
\date{}

\maketitle

\begin{abstract}

We focus on the importance of $q$ moments range used within multifractal detrended fluctuation analysis (MFDFA) to calculate the generalized Hurst exponent spread and multifractal properties of signals. Different orders of detrending polynomials are also discussed.
In particular, we analyze quantitatively the corrections to the spread of generalized Hurst exponent profile $\Delta h$
allowing to extend the previously found by us formulas for large $q$, describing the level of artificial multiscaling in finite signals, to arbitrary narrower range of $q$ moments used in MFDFA technique in distinct applications.

\end{abstract}

$$
$$

The multifractal detrended fluctuation analysis (MFDFA)  \cite{MF-DFA} appears to be nowadays the main tool for investigation of multifractal properties in complex systems and in time series. It has been discussed in more than 500 papers now and applied to complexity issues in variety of topics
(see, e.g.,
\cite{MF-seismology_1}--\cite{czarnecki}).
Very recently, two papers were released on some subtle problems one may encounter in MFDFA. The first article \cite{Drozdz_FENS} indicated the role of detrending polynomials on the final results and pointed at the role of the polynomial order. However, it did not explore wider range of $q$ moments of detrended fluctuations $F_q$ \cite{MF-DFA} recalled below in Eq.(1).

The second paper \cite{arxiv} put an attention on artificial multiscaling effects observed in MFDFA as a result of apparent multifractality caused by the effects of finite length of a signal, i.e., finite size effects (FSE). The  latter phenomenon is significantly magnified when data in series reveal persistency. That paper in turn, discussed only large range of $q$ moments ($-15\leq q\leq 15$) and took into account merely second order polynomials in detrending procedure, while Authors  of Ref.\cite{Drozdz_FENS} considered the smaller range $-4\leq q\leq 4$.
The goal of this article is to make a bridge between semi-analytical formulas obtained in  \cite{arxiv} for the level of artificial multiscaling effects in complex systems calculated for larger $q$'s with the corresponding effects for smaller range of moments $q$, as well as with different detrending polynomials used in MFDFA.

Let us remind that $F_q$, according to the standard definition in MFDFA reads \cite{MF-DFA}

\begin{equation}
F_q(\tau)=\left\{\frac{1}{2N}\sum^{2N}_{k=1} [\hat{F}^2(\tau,k)]^{q/2}\right\}^{1/q}
\end{equation}
where
\begin{equation}
\hat{F}^2(\tau,k)=\frac{1}{\tau}\sum^{\tau}_{j=1}\left\{x_{(k-1)\tau+j}-P_k(j)\right\}^2
\end{equation}
 and $x_j$ ($j=1,\ldots,N\tau$) are data in series, $\tau$ is the size of window box in which detrending is performed, while $P_k(j)$ is the polynomial trend subtracted for $j$-th data in $k$-th window box ($k=1,\ldots,N$).

The power law $F_q(\tau)\sim \tau^{h(q)}$ defines the generalized Hurst exponent $h(q)$ which is crucial within MFDFA to estimate the multifractal properties of a given signal.

Many authors use wider range for $q$ in their calculations and applications (see, e.g.,\cite{zunino,MFDFA_do10_1,zunino1,MFDFA_do10_4}), even $-20\leq q \leq 20$. It particularly concerns problems where scaling is equally good for small and large $q$ values and simultaneously, the generalized Hurst exponent $h(q)$ is well defined monotonic function.
The latter property enables to plot the singularity spectrum $f(\alpha)$ \cite{SFA} as the inverted parabola-like shape and then to read the singularity spectrum spread $\Delta \alpha\equiv\alpha_{max}-\alpha_{min}$ directly from the regular $f(\alpha)$ plot \cite{SFA}. However, if non-monotonic behavior in $h(q)$ is observed \cite{spur-corr-mf}, one cannot built the singularity spectrum $f(\alpha)$ nor to draw any convincing conclusions on multifractality from it, because the Legendre transform linking $h(q)$ with $\alpha(q)$ and $f(\alpha)$ is ambiguous. Also the spread $\Delta h =h(-q)-h(q)$, defined as the difference of generalized Hurst exponents for small negative  and large positive fluctuation moments, is not indicative in this case for $q\rightarrow \infty$  (see, e.g.,\cite{zunino1},\cite{spur-corr-mf}). For instance, in the case of nonstationary data with periodicity, white or color noise added (see, e.g., \cite{spur-corr-mf}) one may see domains where $h(q)$ is either increasing or decreasing with $q$, local maxima in $h(q)$ are formed or even $h(q\rightarrow{-\infty})<h(q\rightarrow{+\infty})$ suggesting that big fluctuations may appear more often than small ones. It is contrary to observations in stationary data \cite{Kantelhardt-arxiv} where it should be the other way round. We will not address such problem in this article, focusing mainly on the influence of moment order $q$ and detrending polynomial order $m$ on the multifractal findings for artificially generated stationary data.

Even if $h(q)$ is a decreasing function of $q$, a few statistically not important small fluctuations may substantially contribute to $F_q$ fluctuation function for moments small enough ($q<0$),  rising the $h(q<0)$ edge of the multifractal spread. An opposite effect occurs for $q>0$ lowering the influence of very large fluctuations. This influence is meaningful for short time series where such accidental fluctuations, not related with multifractal properties of signal, contribute the most. The way to overcome this difficulty is either to restrict calculations to small $|q|$ moments or to calculate fluctuations for larger moments but simultaneously diminishing the effect of very small or large accidental fluctuations taking FSE into account. This way the   initial generalized Hurst exponent spread $\Delta h$ is lowered.  When the first choice is made, there is no clear argument what $|q|$ should be chosen as the maximal range for the considered particular problem, although some light at this issue is shed in recent publication \cite{mexico}. Therefore it seems to be reasonable to have ready to use formulas for corrections due to FSE which are calculated for arbitrary $q$ in a given range. This way the true level of multifractality present in a system can be estimated for given, arbitrary (in some range) value of $q$  when MFDFA technique is applied. We will proceed in this direction in this article.


The strength of multifractality present in data, defined as a spread of generalized Hurst exponent $\Delta h$, should be generally considered as dependent on $q$ parameter range.
Let us introduce a notation
\begin{equation}
\Delta_q h \equiv h(-q) - h(q)
\end{equation}
describing this dependence for any $q\geq0$.
In the case of stationary series, $h(q)$ is a monotonically decreasing function and therefore $\Delta_q h$ increases with $q$ \cite{Kantelhardt-arxiv}.

A number of corrections should be applied to initial results of MFDFA, when the narrower range of $q$ ($q\ll\infty$) is taken for calculations of $\Delta_q h$.
Such corrections can be simply defined as
\begin{equation}
\delta_q h\equiv\Delta_\infty h - \Delta_q h.
\end{equation}

To picture an importance of this dependence, let us first consider a basic model of multifractality, i.e., the generalized binomial cascade model \cite{cascade}.
The generalized Hurst exponent (for $q\neq0$) is described within this model by analytic formula
\begin{equation}
h(q) = \frac{1}{q}\left[1-\left(a^q+(1-a)^q\right)\right]
\end{equation}
where $a$ is a parameter responsible for richness of multifractal properties ($0.5<a<1$).
Eq.(4) enables to determine analytically the spread $\Delta_q h$ at any value of $q$.
In particular, if $q\to\infty$ one obtains
\begin{equation}
\Delta_{\infty} h = \log_2\frac{1-a}{a}.
\end{equation}

Figs. 1a-d reveal numerical results of $\Delta_q h$ dependence, compared with theoretical prediction from Eqs. (4) and (5) shown as a function of the maximal order $q_{max}$ of fluctuation function used in Eq.(1) to extract the $\Delta_q h$ spread.
The order $m$ of detrending polynomial is simultaneously varied in the range $1\leq m\leq7$.
The results shown here generalize findings from Ref.\cite{Drozdz_FENS}, plotted in Fig.5 therein.
Authors of Ref.\cite{Drozdz_FENS} used the singularity spectrum language instead of generalized Hurst exponent (as we did) and found the multifractal features for smaller $q$ range ($-4\leq q \leq 4$).
In our approach, the range up to $q_{max}=20$ was searched through and two different lengths of data were taken into account: $L=2^{16}$ and $L=2^{20}$.

The plots in Figs. 1a-d confirm that $\Delta_q h$ does not depend on the polynomial order up to $m=7$ not only for $q_{max}=4$ but also in much wider range of $q_{max}=20$. This statement is equally valid for short and long data series (compare Figs. 1a-b with Figs.1c-d). One observes also that numerical simulation agrees well with theoretical prediction from binomial cascade models for all $q$ ranges (see Eq. (4)).
In addition, $\Delta_q h$ significantly changes with $q_{max}$.
For series with 'richer' multifractal properties (higher $a$),  $95\%$ of the expected asymptotic multifractal strength $\Delta_\infty h$ is obtained already for $q_{max}=20$, once only $\sim 75\%$ of $\Delta_\infty h$ is reached at $q_{max}=4$ (see Fig. 1b,1d).
For series with lower multifractal content (lower $a$) the situation is worse, since $q_{max}=20$ gives only $80\%$ of $\Delta_\infty h$, while $q_{max}=4$ returns merely $\sim 40\%$ of the value predicted by Eq. (5).

Data generated with stochastic Log-normal and Log-Poisson multiplicative cascade models \cite{stoch_cascades},
shown in Fig. 2a-d, also confirm weak dependence on the order of detrending polynomial.
They also reveal that measurement of $\Delta h$ at $q_{max}=4$ gives only $\sim75\%$ of the $\Delta h$ spread obtained at $q_{max}=15$.

To reduce the influence of accidental fluctuations in short data series on the multifractal findings in signal, one can find this influence in synthetic monofractal data first to reveal the lower threshold (bias) of such phenomena.
Once we turn to monofractal persistent data, the outcomes for multifractal bias resulting from the finite length, i.e., so called FSE multifractal effects \cite{arxiv}, become varying on both: detrending polynomial order $m$ and $q_{max}$.

The spread $\Delta_q h$ for different orders of detrending polynomial $1\leq m \leq 7$ is presented in Figs. 3a-d for artificial fractional Brownian motion signals.
It is clearly visible, that higher detrending polynomial orders ($m>3$) increase the FSE multifractal bias, what supports findings for singularity spectrum $\Delta \alpha$ obtained in Ref.\cite{Drozdz_FENS}, but done for smaller statistics of $10$ series there (see Fig. 2 in Ref.\cite{Drozdz_FENS}). Here we increased this statistcs ten times up to $100$ time series for every $q$ value.
 Fig.3a-d imply that the use of detrending polynomials with order $m>3$ magnifies the multifractal FSE bias in short monofractal signals and therefore, is not recommended in practice.
We will stick to $m=2$ detrending polynomial function in further analysis because it is a safe choice as argued above, being in agreement with the one made in Ref.\cite{arxiv}.

The correction for $q<Q$, ($q>0$) to our previous results calculated at $q_{max}\equiv Q=15$ \cite{arxiv} for $\Delta_q h$ corresponding to maximal FSE bias in persistent series (of mono- or multifractal origin) can be written in similarity with Eq.(4) in the form
\begin{equation}
	\delta_q h(\gamma,L) = \Delta_{Q} h(\gamma,L) - \Delta_q h(\gamma,L)
\end{equation}
where $\delta_q h$ depends obviously on the length of data $L$ and on its long-term memory properties.
The latter property is usually described by the $\gamma$ scaling exponent \cite{gamma_1,gamma_2} of autocovariance function for data increments and is connected with the main Hurst exponent via relation $\gamma=2-2H$ \cite{rel_gamma_H}.

The results of Figs. 3a-d can be also shown in complementary Figs.4a-i, indicating dependence $\delta_q h$ versus $q$ for variety of $\gamma$'s and time series lengths.
All plots in Figs.4a-i clearly confirm the existence of a threshold $q_T \sim 4\div6$ (lower threshold value applies to more persistent series) dividing $q$ range into two domains of different $\delta_q h$  behavior.
One notices a linear dependence between $\delta_q h$ and $q$ for $q>q_T$, what enables to write a simple relation in this range of $q$
\begin{equation}
\delta_q h(\gamma,L) = A(\gamma,L)\left(Q-q\right).
\end{equation}
The slope $A$ depends only on the persistency level $\gamma$ in data and on time series length $L$.
This relation  fully describes the nature of correction for the multifractal profile spread.

Let us plot first the $A(\gamma, L)$ values against the scaling exponent $\gamma$  as in Fig 5.
The results for just three distinct signal lengths ($L=2^{12}, L=2^{16}$ and $L=2^{20}$) are shown here but very similar outcomes were found by us for remaining lengths as well.
One concludes that
\begin{equation}
A(\gamma,L) = A_\gamma(L)\gamma+B_\gamma(L)
\end{equation}
where the coefficients $A_\gamma$ and $B_\gamma$ may vary only with the signal length.
The further analysis of $A_\gamma(L)$ and $B_\gamma(L)$ is presented in Fig.6.
First, it shows a power law decay of  $B_\gamma(L)$ with $L$
\begin{equation}
B_\gamma(L) = aL^{-\mu}.
\end{equation}
Then, since $A_\gamma(L)$ is almost constant and negligibly small $(|A_\gamma(L)|<3\times10^{-3}$, and furthermore multiplied by $0<\gamma<1$ when entering Eq.(9)) for all signal lengths in comparison with $B_\gamma\sim 10^{-2}$, the main contribution to $\delta_q h(\gamma,L)$ can be assumed to come entirely from $B_\gamma$ term.
Therefore, $\delta_q h(\gamma,L)$ finally reads
\begin{equation}
	\delta_q h(\gamma,L) = aL^{-\mu}(Q-q)
\end{equation}
where $a=0.019\pm 10^{-3}$ and $\mu=0.084\pm0.003$ are found at $Q=15$  from the fit to central values in Fig.(6).

The latter formula expands the usefulness of semi-analytical relations obtained  for the multifractal FSE bias level in Ref.\cite{arxiv} and enables to use them for signals investigated within MFDFA also when much more narrower range of moment order $4\leq |q| <Q$ is used. The combined formula for the multifractal FSE bias at arbitrary $4\leq |q|\leq Q$ reads therefore (see \cite{arxiv})
\begin{equation}
\Delta_q h(\gamma,L)=C_1 L^{-\eta_1}\gamma + C_0 L^{-\eta_0} (1-\gamma) -  aL^{-\mu}(Q-q)
\end{equation}
where
\begin{equation}
\Delta_Q h(\gamma, L) \equiv C_1 L^{-\eta_1}\gamma + C_0 L^{-\eta_0} (1-\gamma)
\end{equation}
was found in \cite{arxiv} with the numerical estimation of constants $C_0$, $\eta_0$, $ C_1$, $\eta_1$ for $Q=15$.
These results will be quantitatively similar for detrending polynomial orders $m<5$, as shown in Fig.(3a-d).
The quantitative corrections to the observed multifractal bias effects given by Eq.(12) are visualized in Fig. 7 for four chosen lengths of moderately persistant signals ($\gamma=0.5$).

Thanks to the formula in Eq.(12) one is able to compare results of multifractal spread in MFDFA obtained at different values of $q_{max}$. More precisely, if we get two distinct results of initial (naked) multifractal spread, say $\bar{\Delta} h_{q_1}$ and $\bar{\Delta} h_{q_2}$, calculated for $q_1>q_2>0$, then the real content of multifractality cannot be estimated from such spreads alone unless the unbiased spreads
 $\bar{\Delta} h_{q_1}-\Delta_{q_1} h(\gamma,L)$ and  $\bar{\Delta} h_{q_2}-\Delta_{q_2} h(\gamma,L)$ are considered which take into account corrections from Eq.(12). If the latter two unbiased spreads are nearly the same, one can conclude that $q_2$ moment order is sufficient to reveal the influence of all small (large) fluctuations on the multifractal properties of a system. Otherwise, there is a need to consider even higher $q$ moments since it is still possible to find not accidental and statistically important fluctuations amending the multiscaling behavior of  such complexity.

This phenomena can be illustrated with examples of different parts of financial data. A number of observed $\bar{\Delta}h_q$ spreads is compared with their unbiased partners $\bar{\Delta} h_q-\Delta_q h$ and shown in Fig.8a-d. We analysed examples of short parts ($L=10^3$), medium part ($L=6\times10^3$) and all historical closure daily data ($L\sim 15000$) from S\&P500 index \cite{yahoo}.
It is seen that the naked biased multifractal spread $\bar{\Delta}h_q$ continuously grows  with $q$ in all cases, while the unbiased spread calculated as a difference of naked spread and the multifractal bias taken from Eq.(12) tends asymptotically to some constant value.
Only the latter one describes the real multifractal content of the searched financial signal.
In all presented cases, the
asymptotic value of unbiased multifractal spread is reached at $q_{max}\approx15$.
The unbiased multifractal spread at lower $q$ ($q_{max}\lesssim4$) is expected to be much smaller (see Fig.8 in conjunction with Fig.4) than the asymptotic value at higher $q$ arguing for higher $q$ moments as better choice here.
The short $q$ range may even lead to observed multifractal spread below the FSE threshold (see, e.g., Fig.8b).


These examples ground the role of proposed analysis  for estimation of true multifractal features in arbitrary complex systems.
\begin{figure}[p]
\centering
	\subfloat[][]{ \includegraphics[width=8truecm]{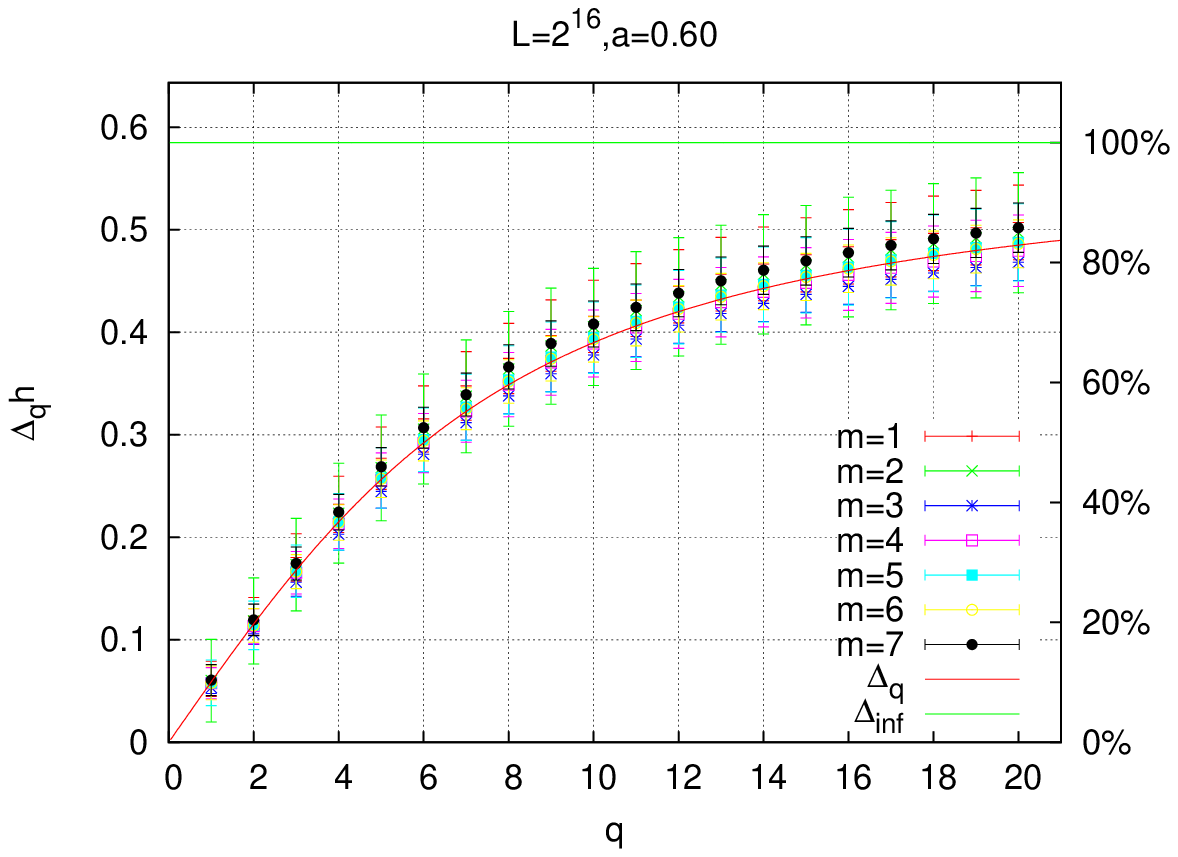} }
	\subfloat[][]{ \includegraphics[width=8truecm]{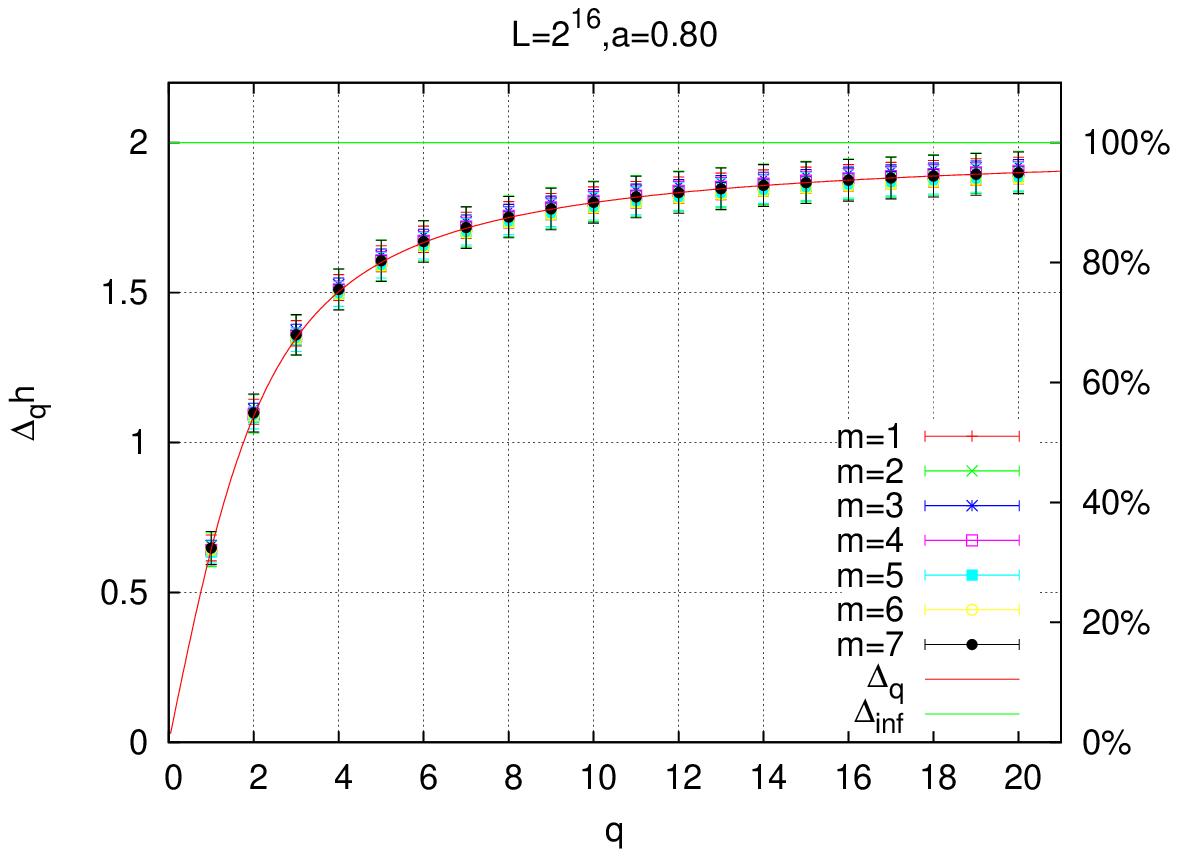} }

	\subfloat[][]{ \includegraphics[width=8truecm]{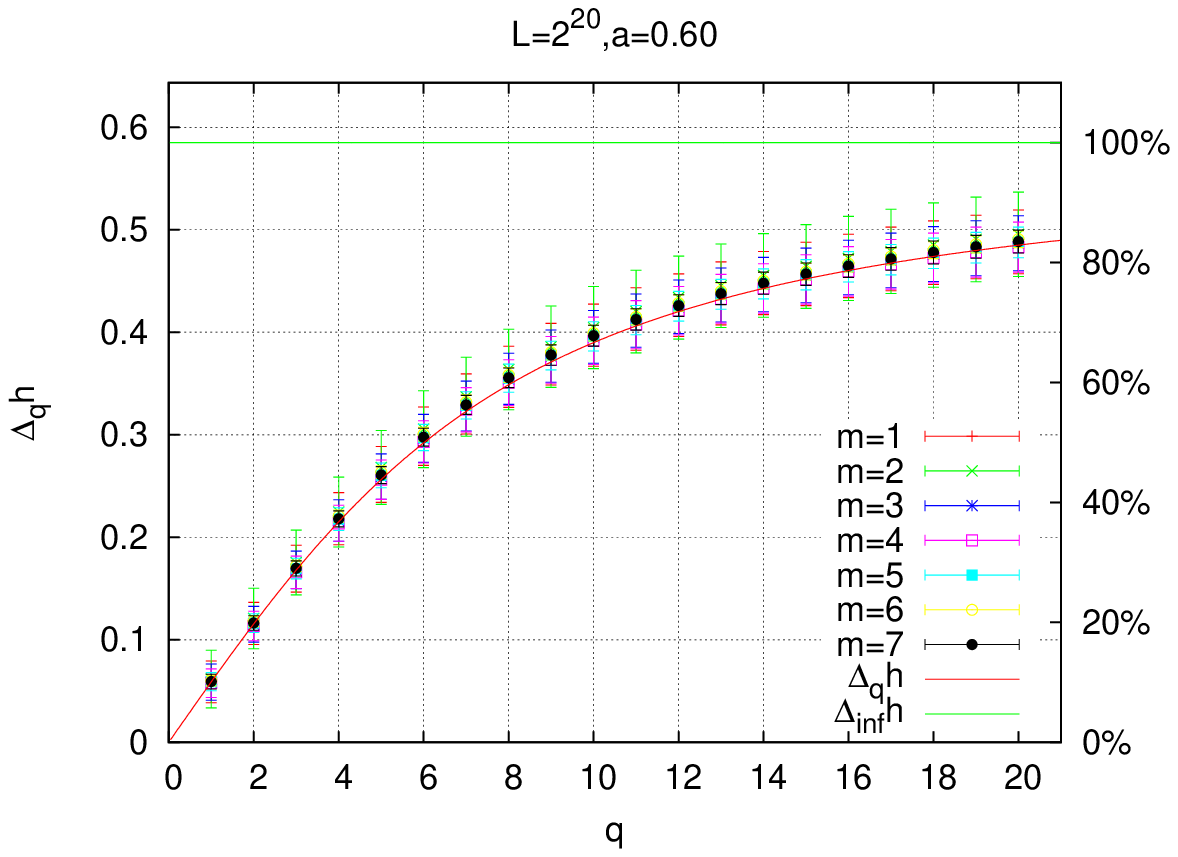} }
	\subfloat[][]{ \includegraphics[width=8truecm]{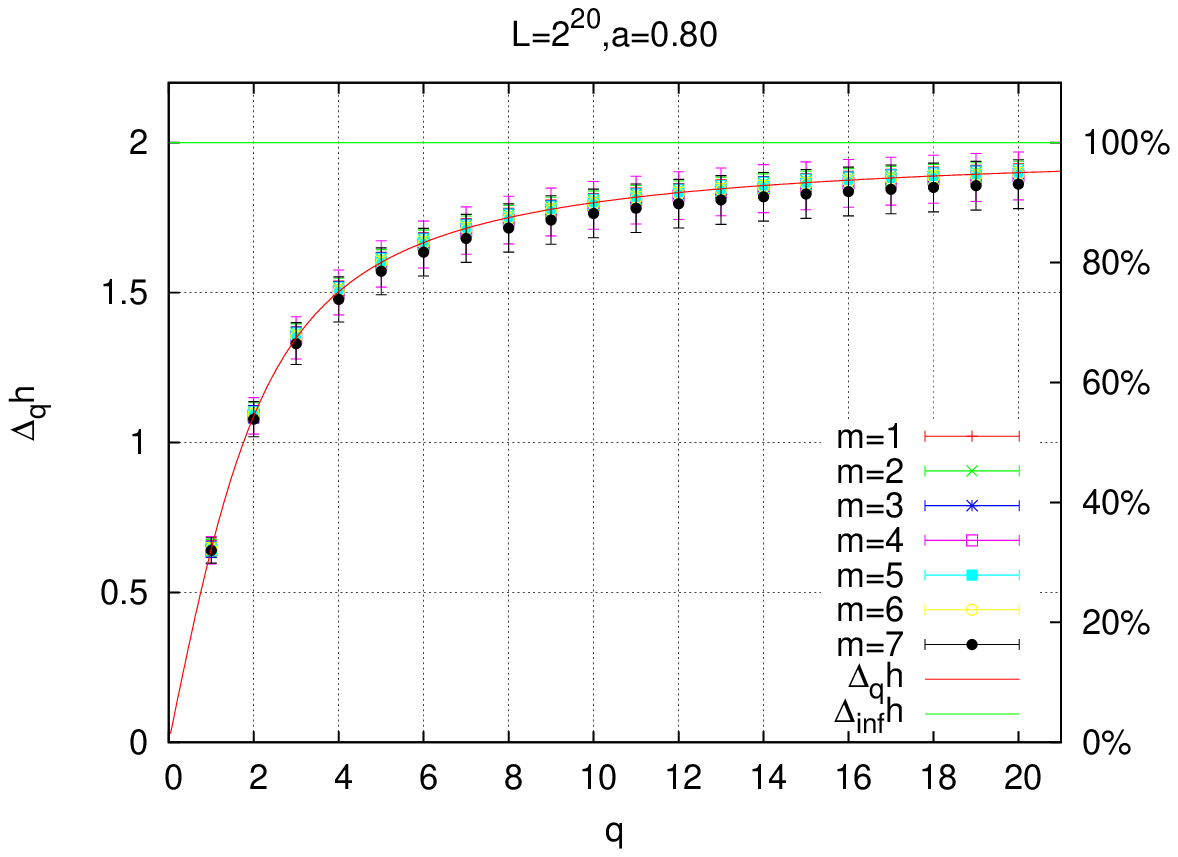} }
\caption{
Dependence of the multifractal profile spread $\Delta_q h$ on the range of $q$ moments used for its evaluation. The results are obtained for synthetic data generated with binomial cascade model for
two lengths $L=2^{16}, 2^{20}$ and two cascade parameter values: $a=0.60, 0.80$.
Results for various orders of detrending polynomial ($m=1,2,\ldots,7$) are plotted.
The corresponding theoretical prediction for $\Delta_q h$ (red curve) and $\Delta_\infty h$ (green line) are also presented.
The uncertainties visible in this figure were calculated on an ensemble of $10^2$ independent realizations.
}
\label{cascades}
\end{figure}

\begin{figure}[p]
\centering
	\subfloat[][]{ \includegraphics[width=8truecm]{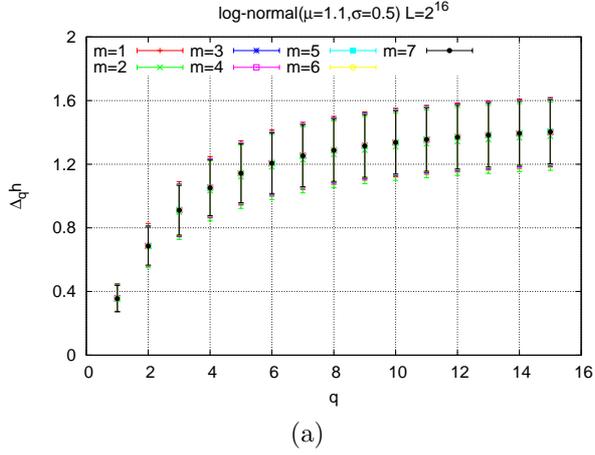} }
	\subfloat[][]{ \includegraphics[width=8truecm]{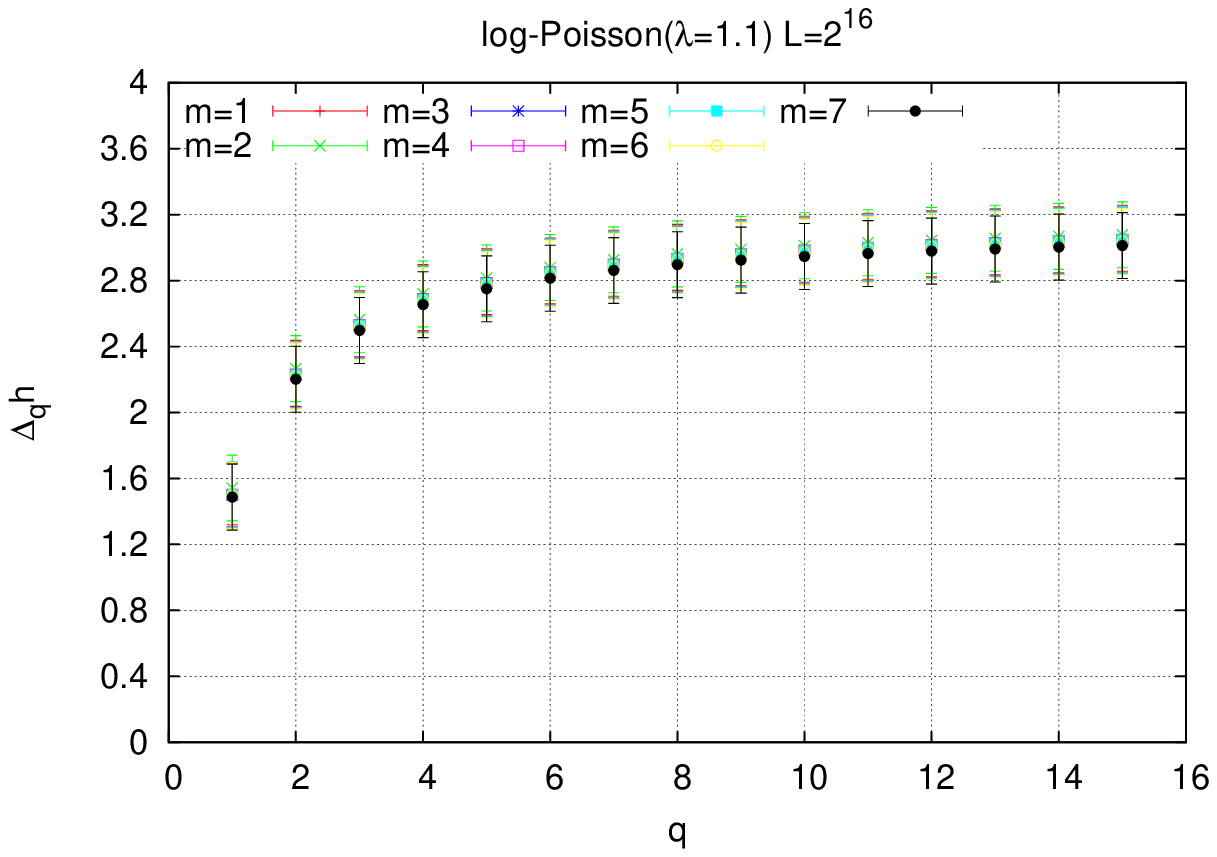} }

	\subfloat[][]{ \includegraphics[width=8truecm]{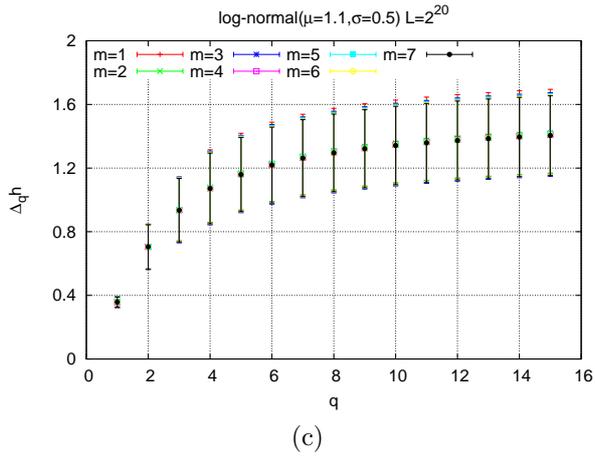} }
	\subfloat[][]{ \includegraphics[width=8truecm]{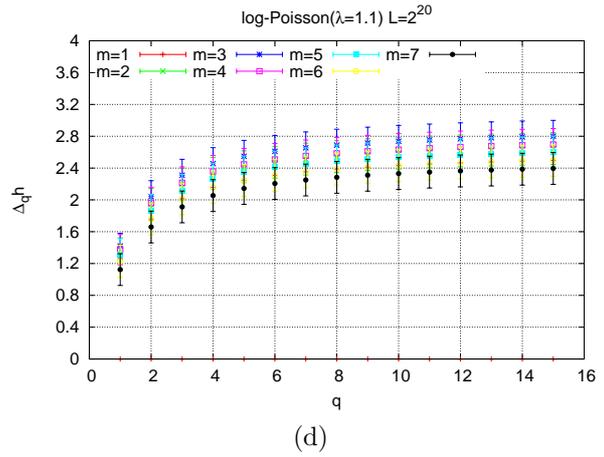} }
\caption{
The same as in Fig.1, but calculated for data generated with stochastic multiplicative cascade model.
Two lengths $L=2^{16}, 2^{20}$ and two distributions (Log-normal in (a),(c) and Log-Poisson in (b),(d)) are considered. Numbers in parenthesis describe parameters of distribution (mean, standard deviation).
}
\label{dh_ffm}
\end{figure}
\begin{figure}[p]
\centering
	\subfloat[][]{ \includegraphics[width=8truecm]{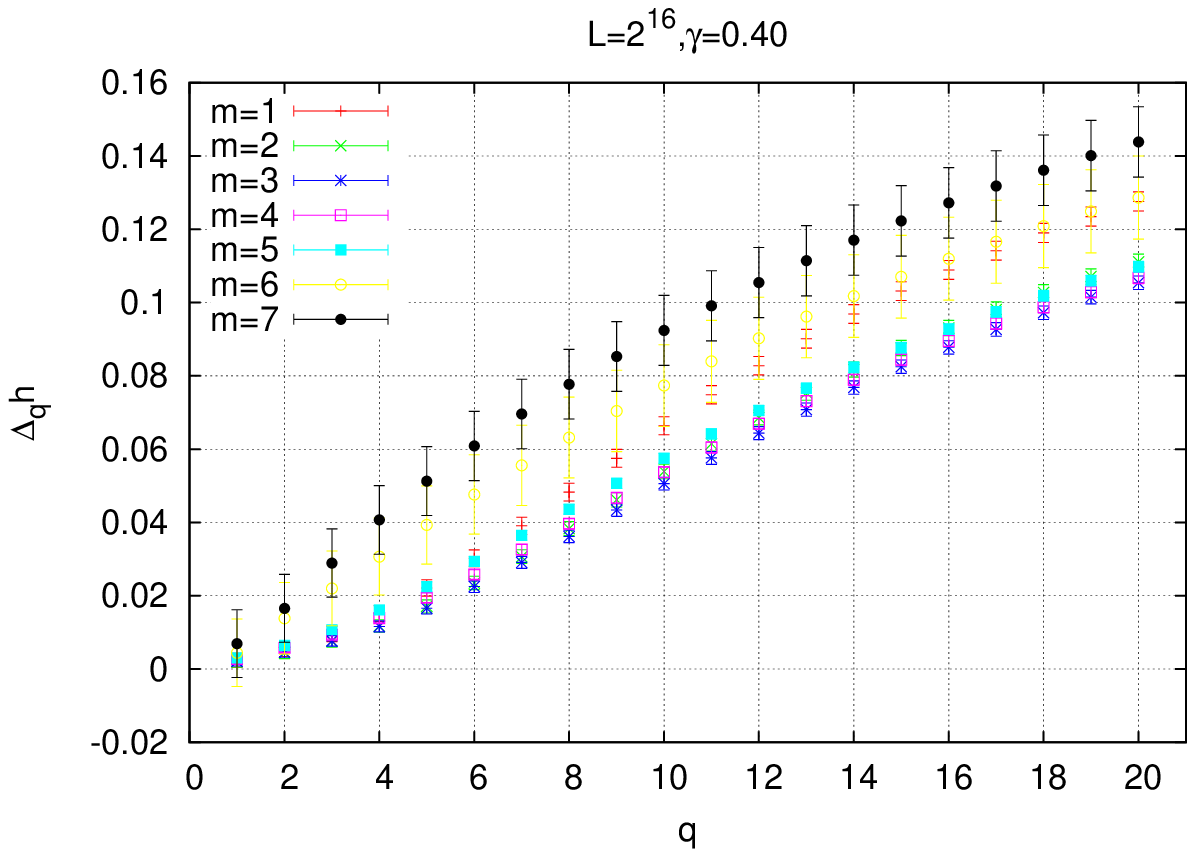} }
	\subfloat[][]{ \includegraphics[width=8truecm]{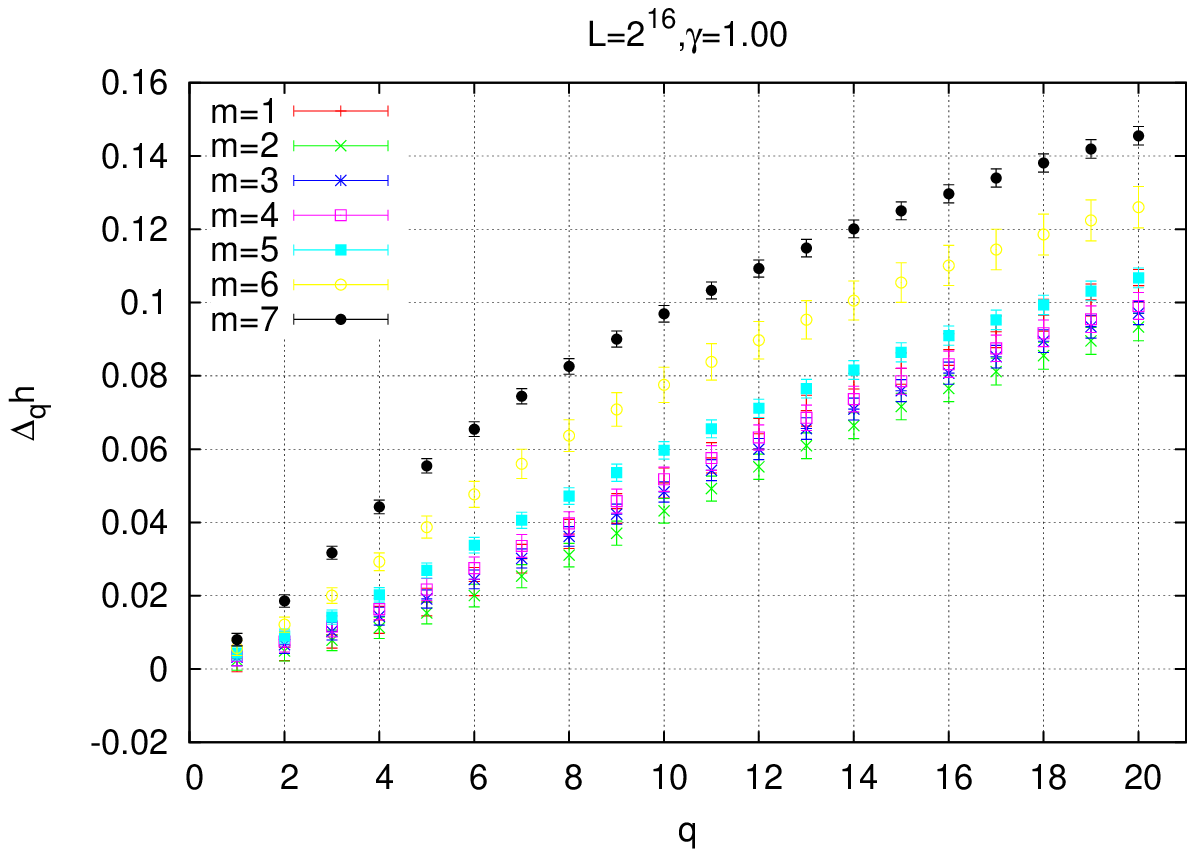} }

	\subfloat[][]{ \includegraphics[width=8truecm]{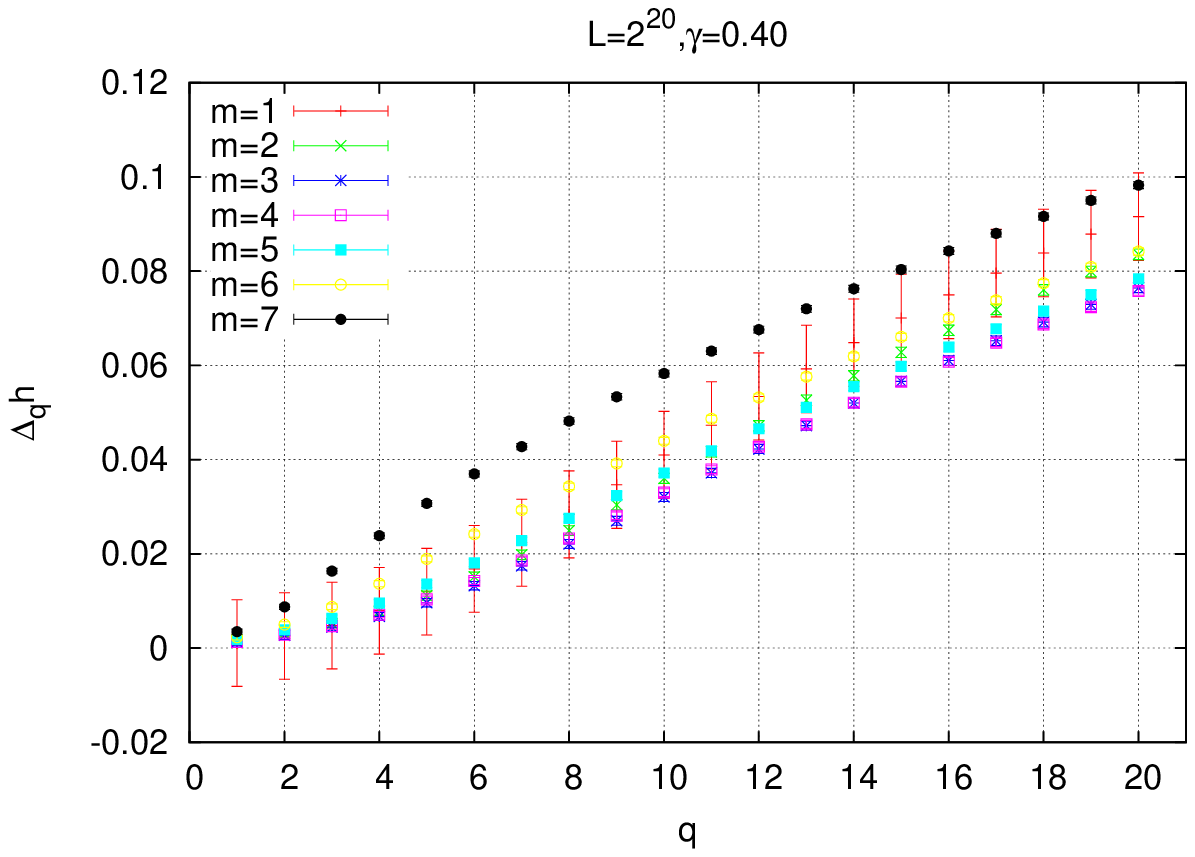} }
	\subfloat[][]{ \includegraphics[width=8truecm]{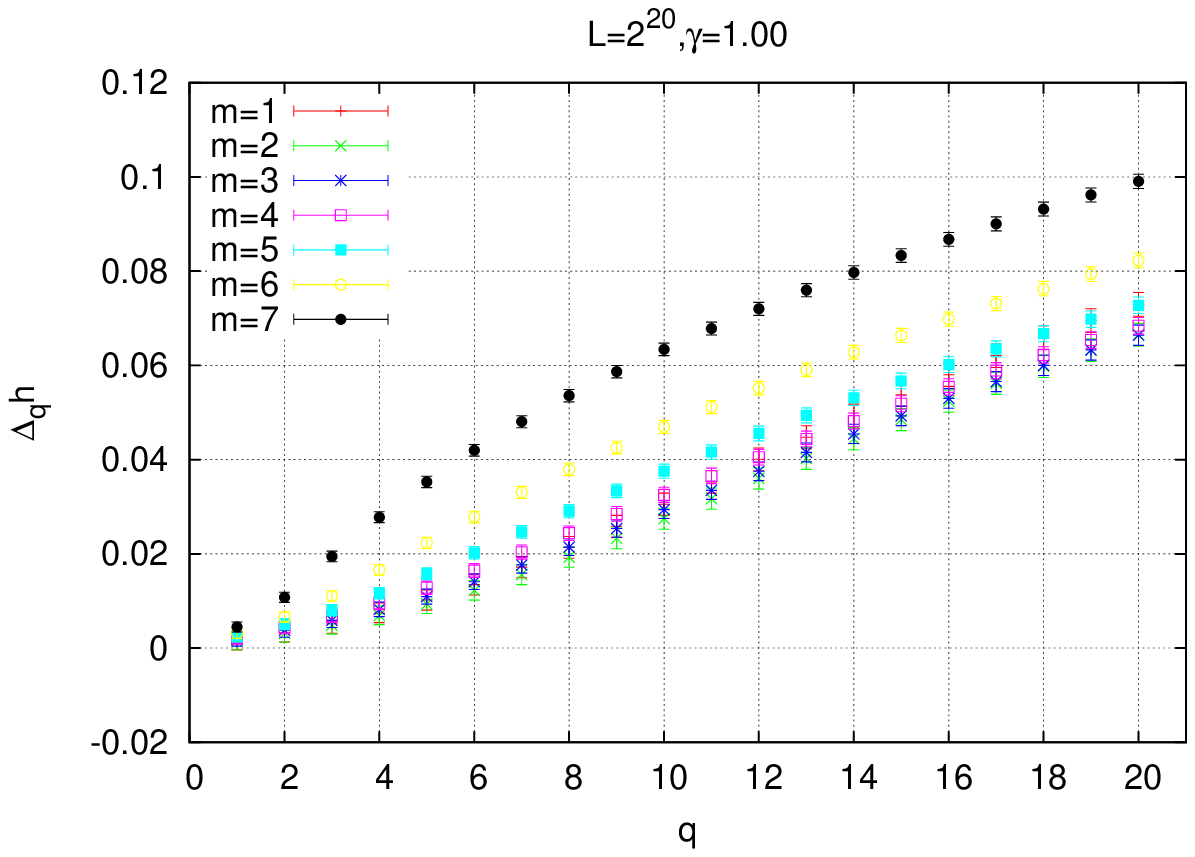} }
\caption{
The same as in Fig.1, but calculated for monofractal data generated in FFM algorithm.
Results are shown for two lengths ($L=2^{16}, 2^{20}$), persistent ($\gamma=0.4$) or uncorrelated $\gamma=1.0$) data, and for several orders of detrending polynomials.
}
\label{dh_ffm}
\end{figure}
\begin{figure}[p]
\centering
	\subfloat[][]{ \includegraphics[width=5truecm]{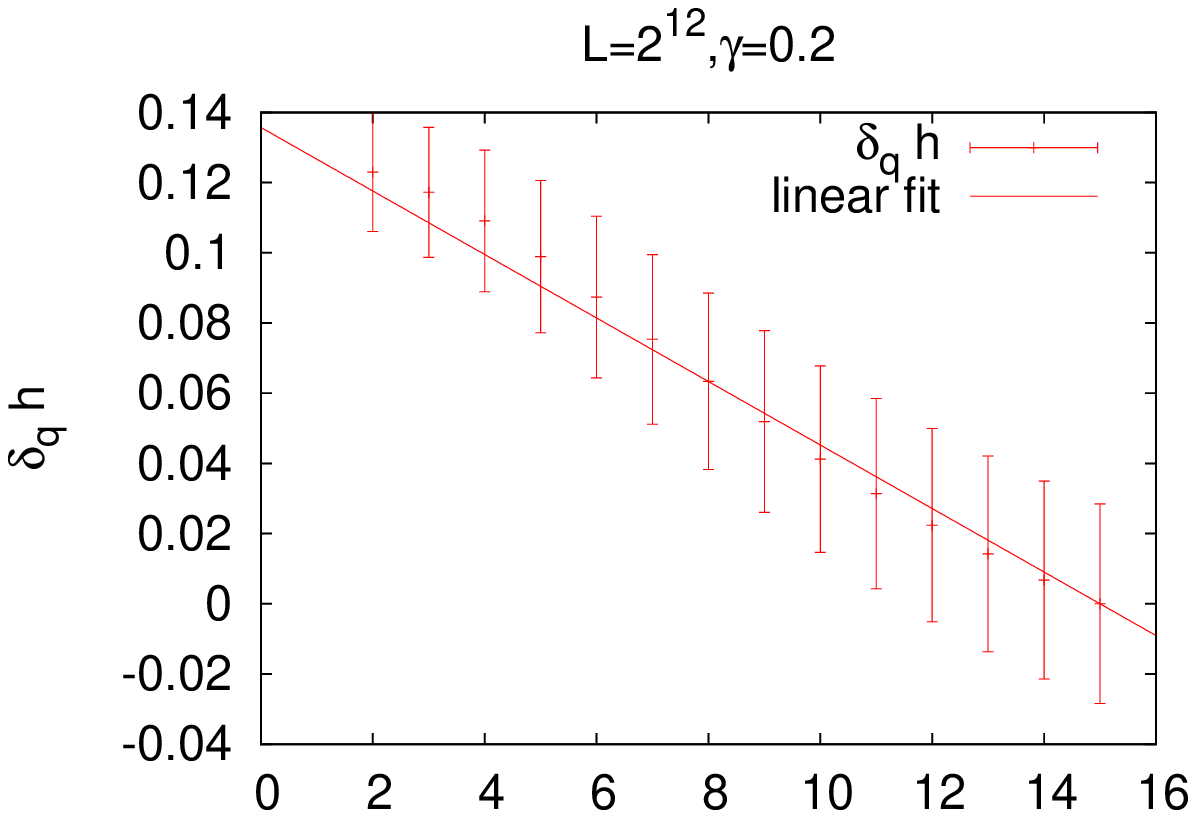} }
	\subfloat[][]{ \includegraphics[width=5truecm]{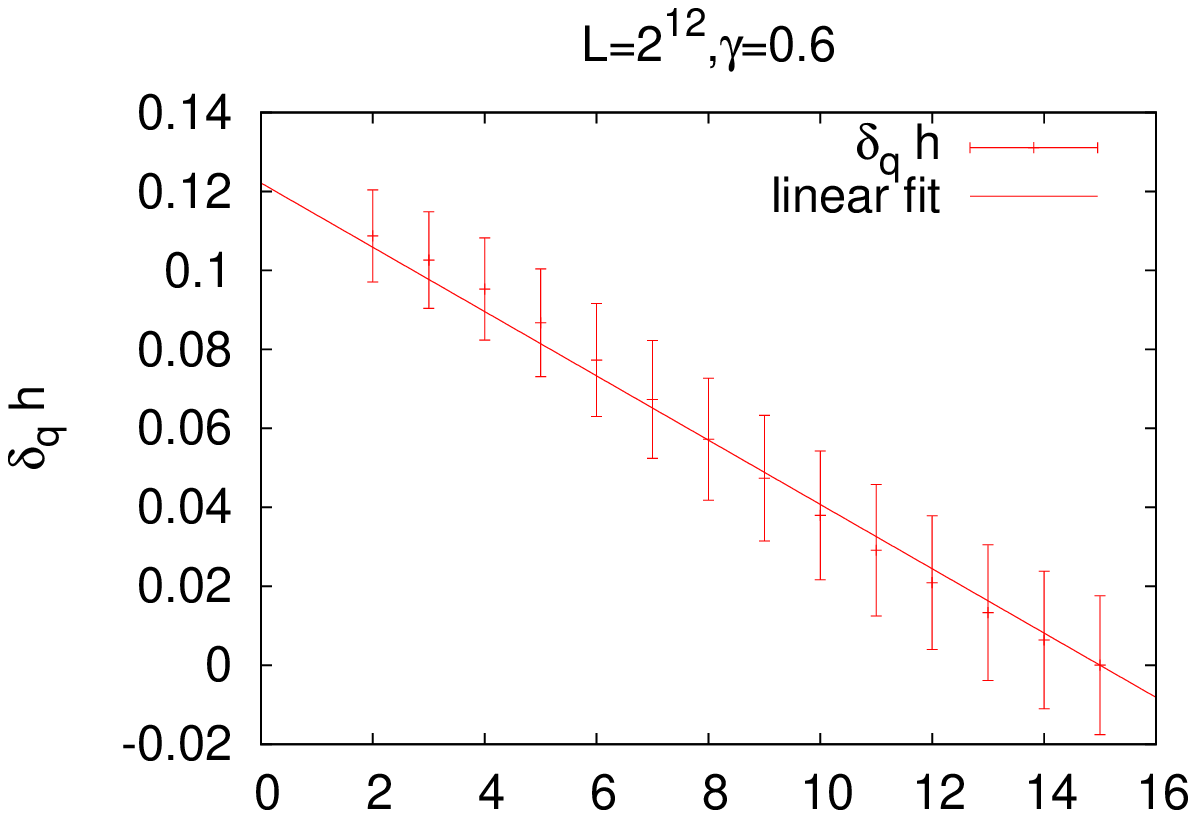} }
	\subfloat[][]{ \includegraphics[width=5truecm]{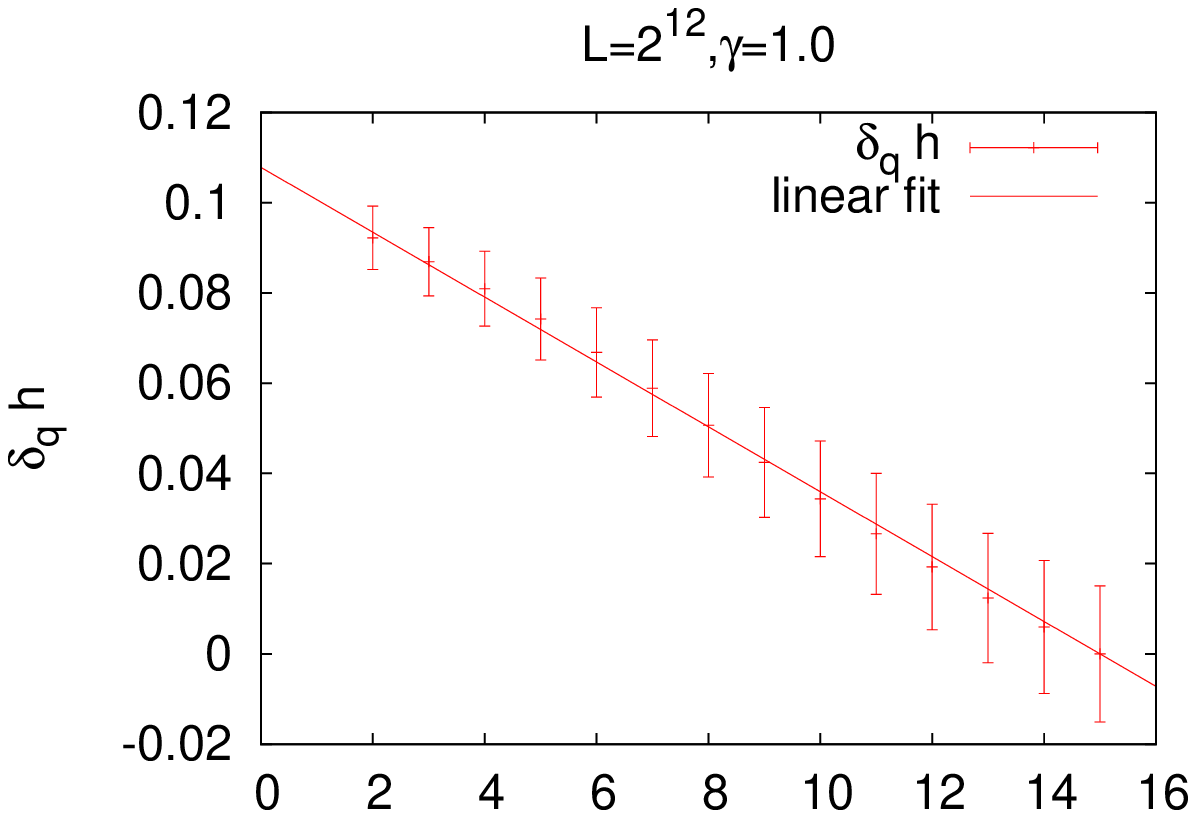} }

	\subfloat[][]{ \includegraphics[width=5truecm]{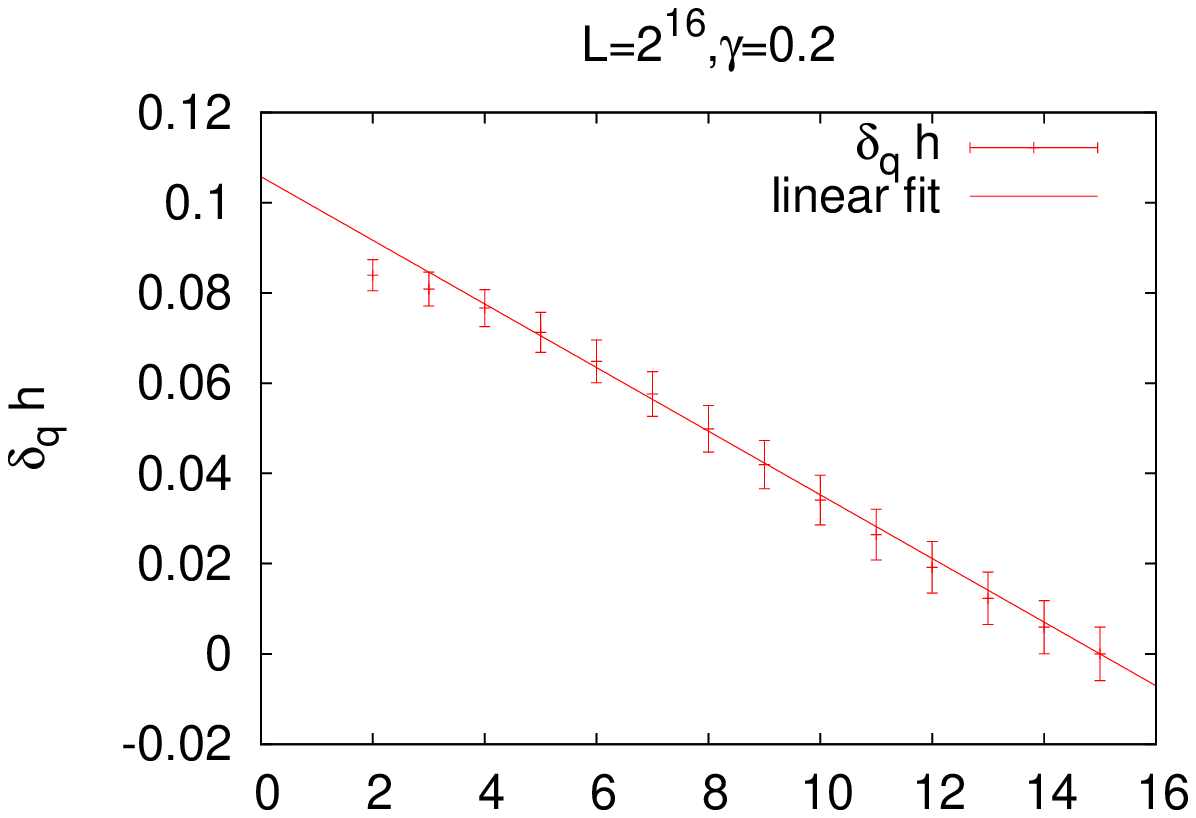} }
	\subfloat[][]{ \includegraphics[width=5truecm]{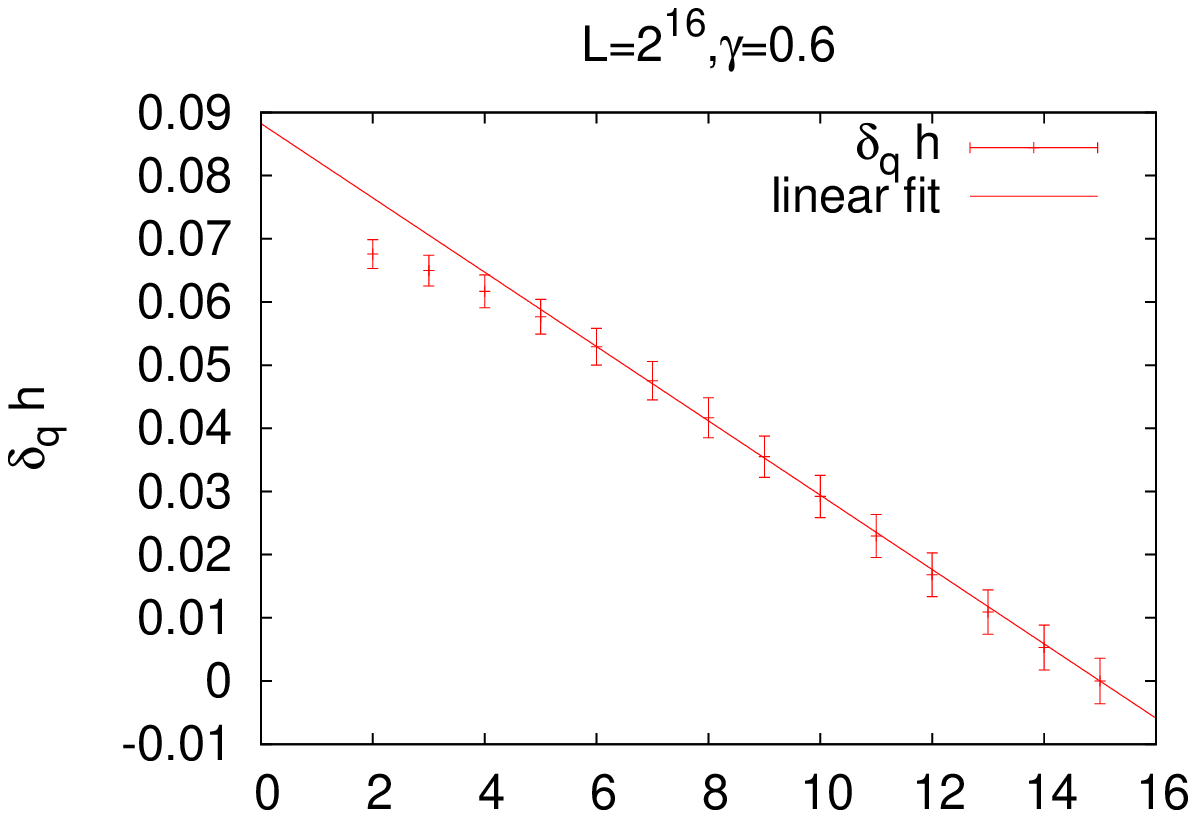} }
	\subfloat[][]{ \includegraphics[width=5truecm]{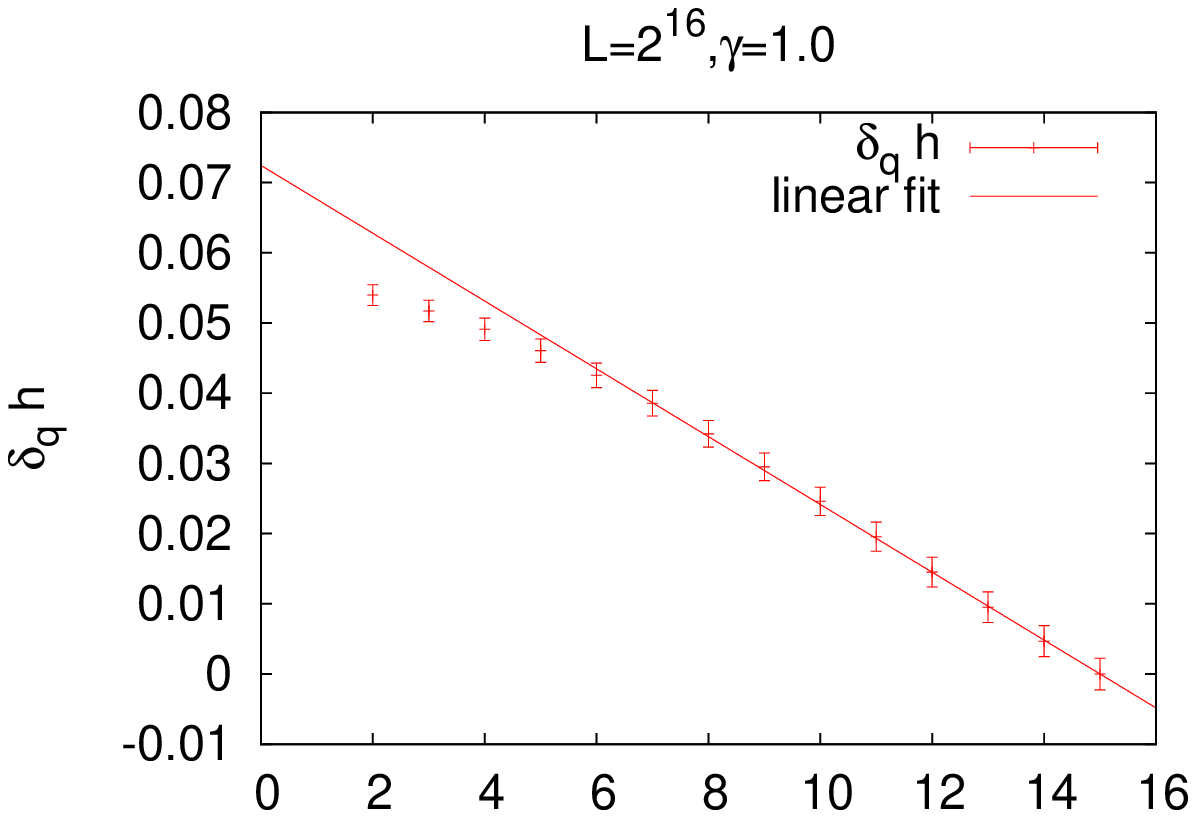} }

	\subfloat[][]{ \includegraphics[width=5truecm]{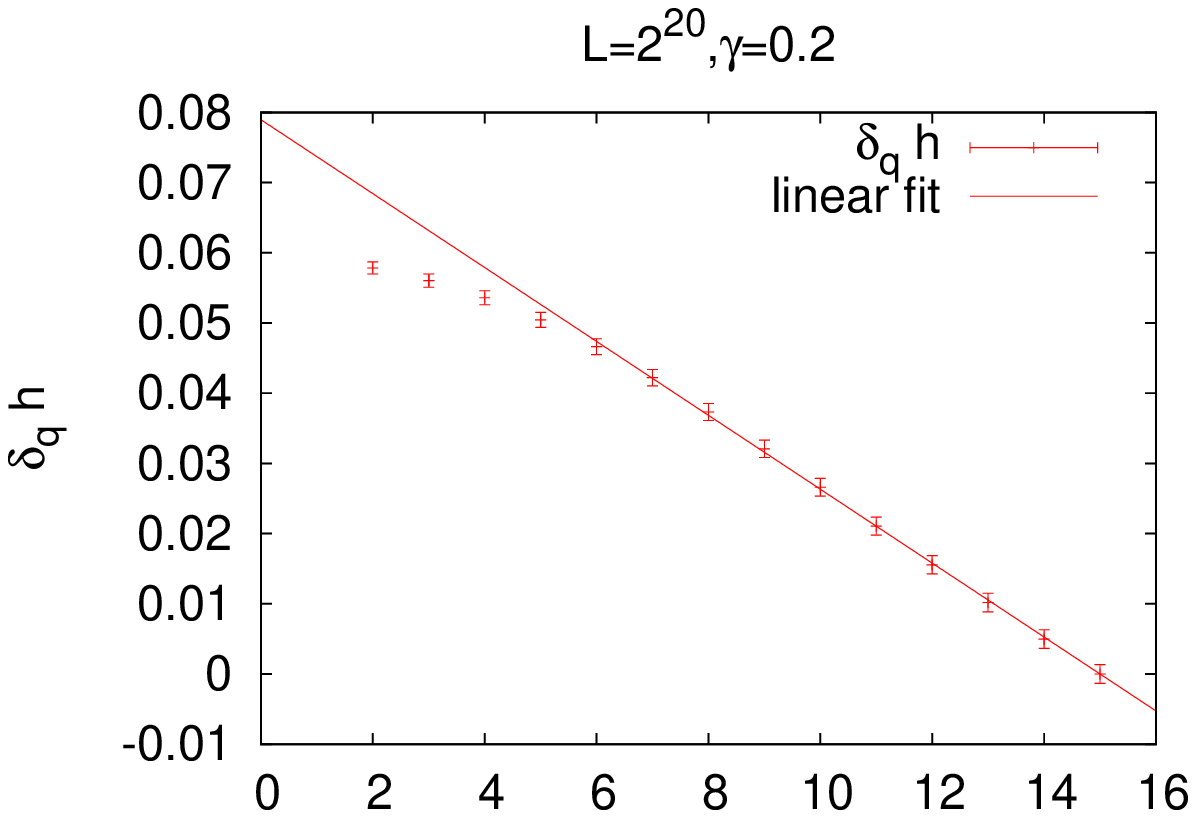} }
	\subfloat[][]{ \includegraphics[width=5truecm]{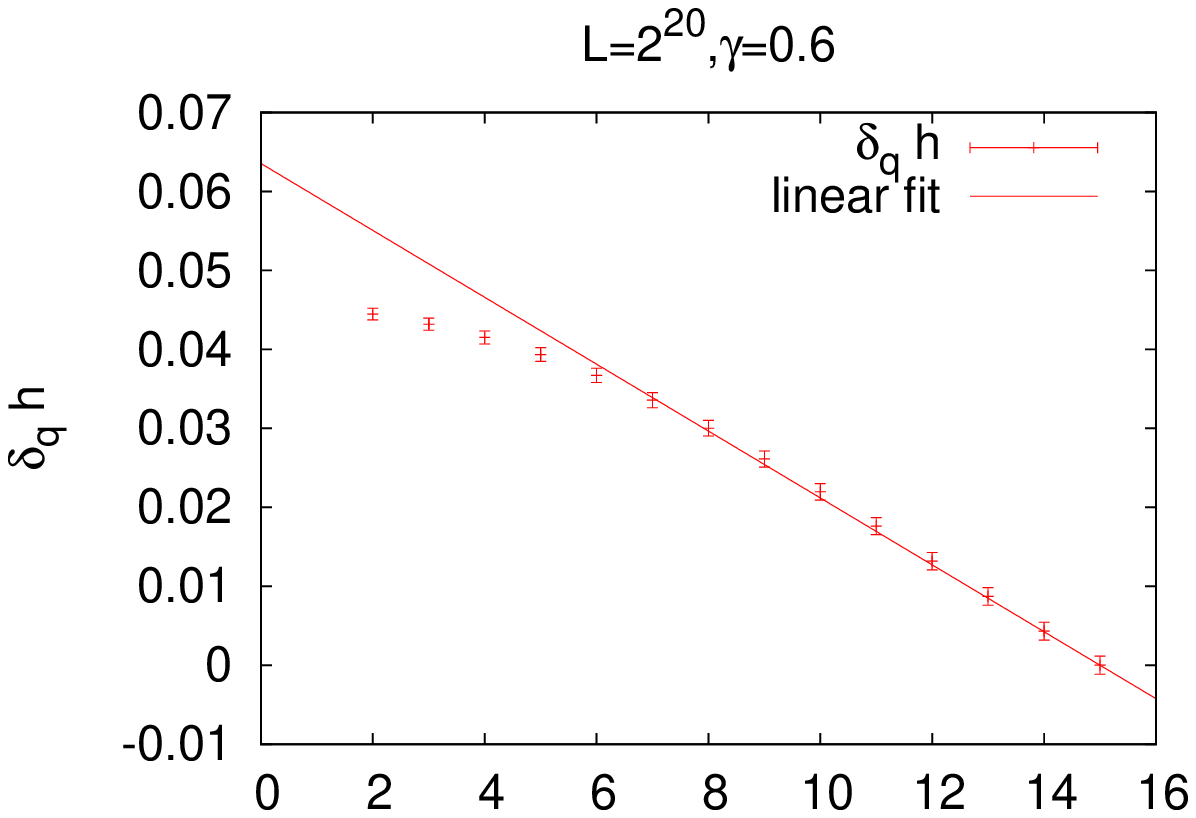} }
	\subfloat[][]{ \includegraphics[width=5truecm]{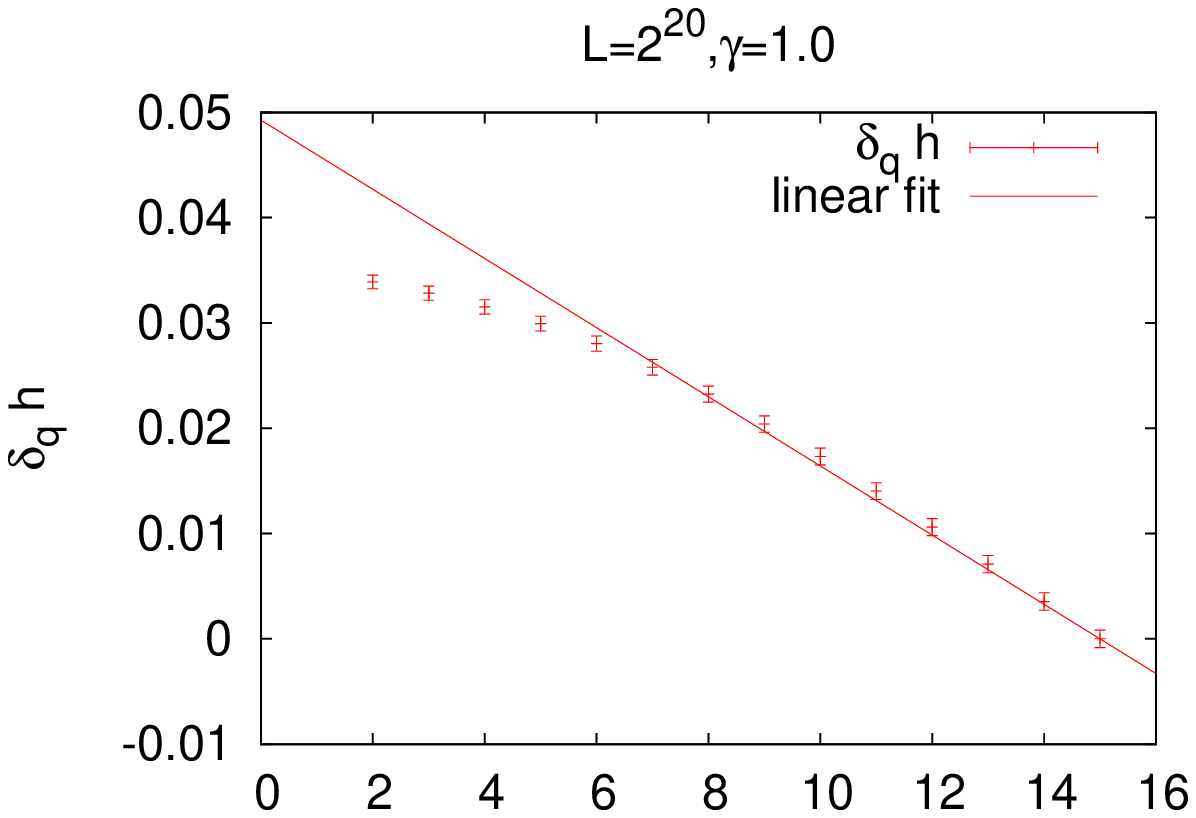} }
\caption{
Correction $\delta_q h$ versus $q$ (see Eq.(7)) gathered for three lengths $L=2^{12},2^{16},2^{20}$ and three persistency levels $\gamma=0.2,0.6,1.0$.
A threshold value $q_T=4\div6$ is seen above which linear dependence is evident.
The uncertainties were calculated on an ensemble of $10^2$ independent realizations.
}
\label{corr_ffm}
\end{figure}

\begin{figure}[p]
\centering
\includegraphics[width=11truecm]{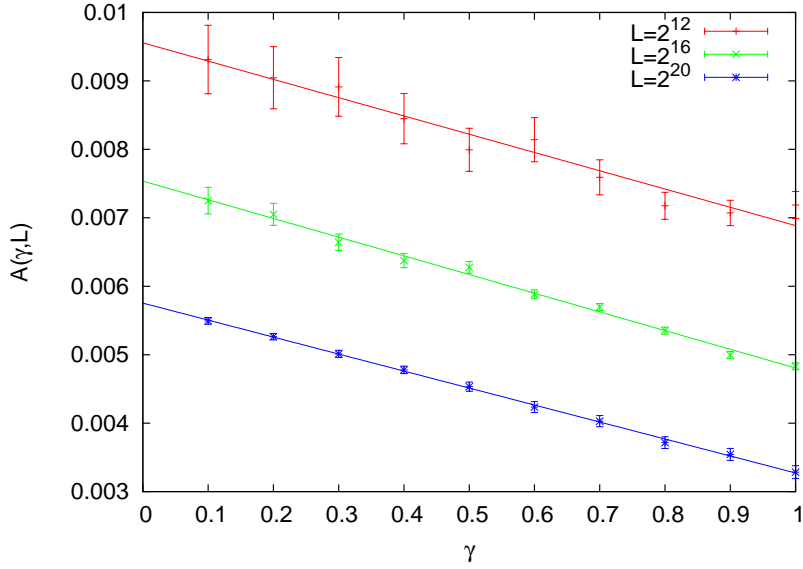}
\caption{
Dependence of linear coefficient $A(\gamma, L)$ from Eq.(8) on long-range memory parameter $\gamma$.
The fitted linear dependence is described by Eq.(9).
The uncertainties arise from the statistics of $10^2$ independent realizations.
}
\end{figure}
\begin{figure}[p]
\centering
\includegraphics[width=12truecm]{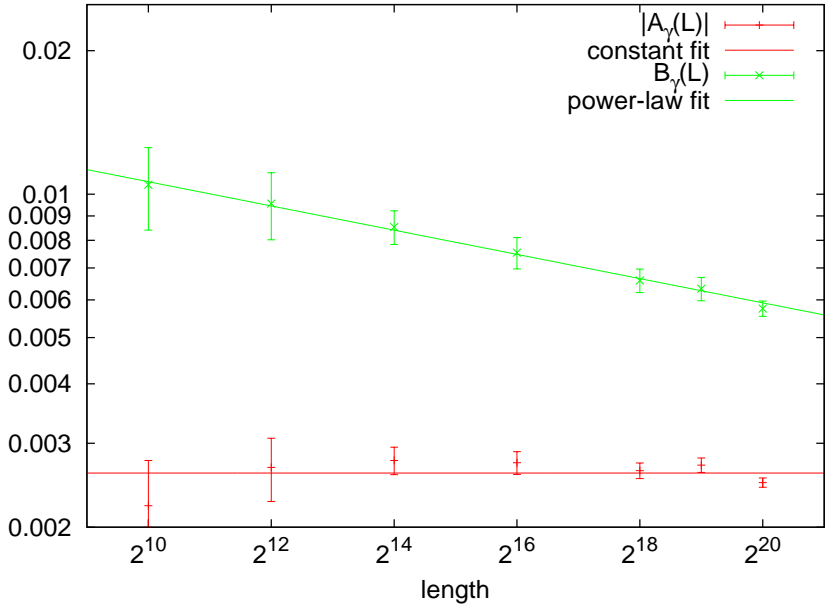}
\label{length_fit}
\caption{
Dependence of coefficients $|A_\gamma|$ and $B_\gamma$ from Eq. (9) on the length of analysed series.
A power-law dependence on $L$ is found for free parameter $B_\gamma$.
Statistics of $10^2$ independent realizations applies.
}
\end{figure}

\begin{figure}[p]
\centering
\includegraphics[width=12truecm]{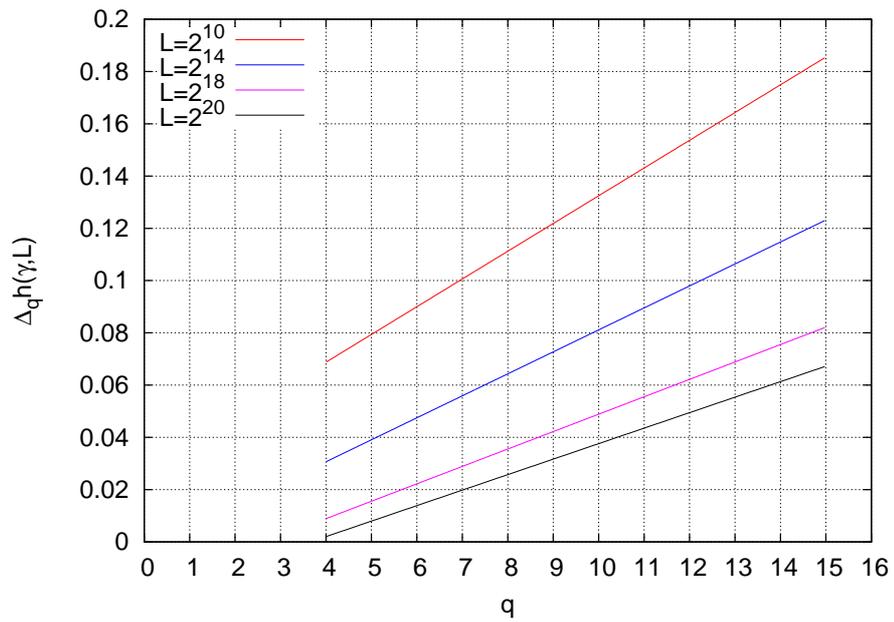}
\caption{
Threshold levels of artificial multifractality in MFDFA caused by  accidental fluctuations contributing to FSE for different values of $q$ moments. Exemplary plots are shown for several lengths of moderately persistent  signals ($\gamma=0.5$).
}
\label{simulated}
\end{figure}
\begin{figure}[p]
\centering
	\subfloat[][]{ \includegraphics[width=8truecm]{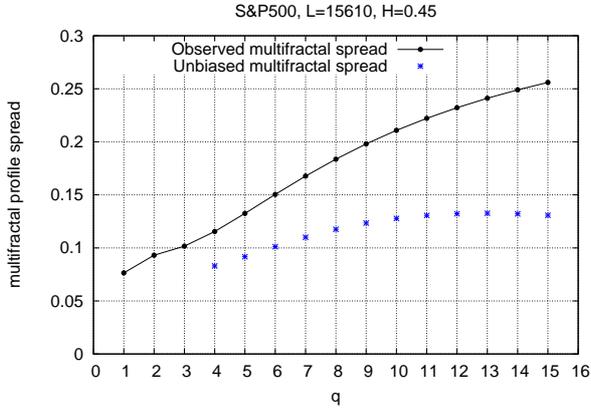} }
	\subfloat[][]{ \includegraphics[width=8truecm]{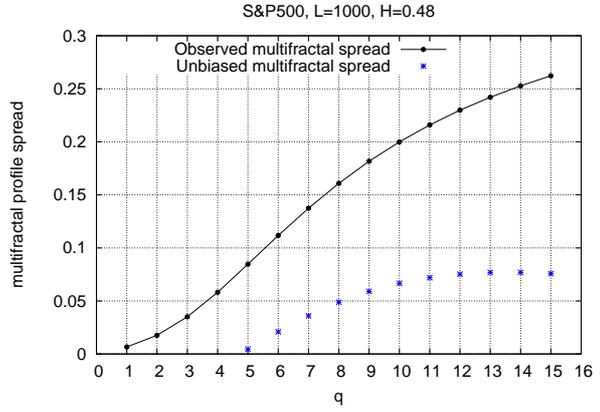} }

	\subfloat[][]{ \includegraphics[width=8truecm]{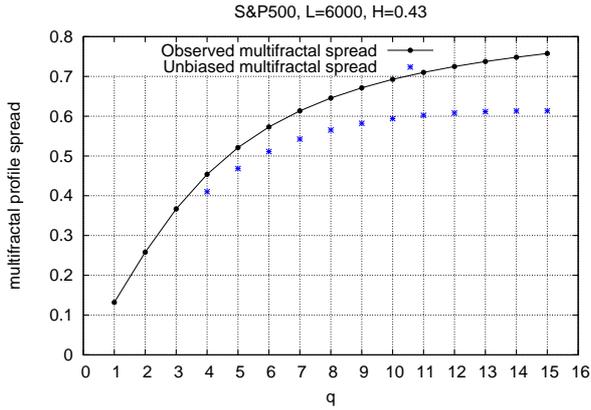} }
	\subfloat[][]{ \includegraphics[width=8truecm]{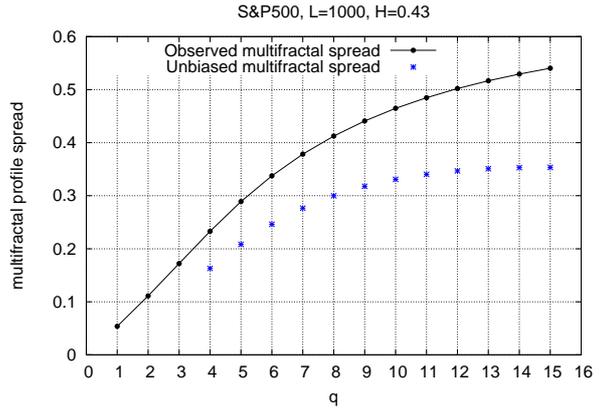} }
\caption{
Comparison of biased $\bar{\Delta}h_q$ (black) and unbiased $\bar{\Delta} h_q - \Delta_q h$ (blue) multifractal spreads for historical closure data of S\&P500 index:
a) since 3/01/1950 till 13/01/2012; b) since 31/12/1969 till 14/12/1973; c) since 14/12/1973 till 10/09/1997; d) since 10/09/1997 till 13/01/2012.
The asymptotic behavior of unbiased multifractal spread at high $q$ values is confirmed for all lengths and for various periods in presented cases of financial data.
}
\label{dh_ffm}
\end{figure}

\end{document}